# Nanolithography based on real-time electrically-controlled indentation with an atomic force microscope for nanocontacts elaboration

K. Bouzehouane[a)]

Unité Mixte de Physique CNRS/Thales, Domaine de Corbeville, 91404 Orsay Cedex, France

S. Fusil

Université d'Evry, Bat. des Sciences, rue du père Jarlan, 91025 Evry Cedex, France

M. Bibes, J. Carrey[b)], T. Blon[c)], P. Seneor and V. Cros

Unité Mixte de Physique CNRS/Thales, Domaine de Corbeville, 91404 Orsay Cedex, France

L.Vila

Unité Mixte de Physique CNRS/Thales, Domaine de Corbeville, 91404 Orsay Cedex, France and Laboratoire PCPM, Université Louvain-La-Neuve, Place croix du sud, 1348 Louvain-La-Neuve, Belgium

Abstract

We report on the fabrication of nanocontacts by indentation of an ultrathin insulating photoresist layer deposited on various types of conductive structures. A modified atomic force microscope (AFM) designed for local resistance measurements is used as a nanoindenter. The nanoindentation is performed while measuring continuously the resistance between the conductive tip of the AFM and the conductive layer, which is used as the trigger parameter to stop the indentation. This allows an extremely accurate control of the indentation process. The indented hole is subsequently filled by a metal to create a contact on the underlying layer. We show that nanocontacts in the range of 1 to 10 $nm^2$ can be created with this technique.

[a)] corresponding author (karim.bouzehouane@thalesgroup.com)

[b)] present address : Center for Magnetic Recording Research, University of California San Diego, La Jolla CA 92093-0401, USA

[c)] present address : Centre d'Elaboration de Matériaux et d'Etudes Structurales, 17 rue Jeanne Marvig, 31400 Toulouse, France.

The elaboration of nano-objects and nanocontacts is expected to show new insight in many fields of physics. They allow for instance to study ballistic transport, conductance quantization [1], inelastic scattering [2,3,4], magnetization reversal by spin injection [5] and electron injection into the quantized levels of small clusters [6,7]. Ferromagnetic nanocontacts obtained by the break junction method [8], by electrodeposition [9] or using tips [10] also yield interesting effects of magnetoresistance. Besides, nanometric contacts could be used as a technological trick to fabricate devices on materials exhibiting a high heterogeneity of their surface properties at the micrometer scale. This is the case when one wants to define tunnel junctions from heterostructures with non-perfect interfaces (containing pinholes for instance). The main difficulty to fabricate nanocontacts is a lack of precise control of their dimensions and electrical properties at the nanometer scale, often leading to unstable contacts with hardly reproducible results. We recently developed a technique [11] to create artificial nanocontacts by a combination of force-controlled indentation of an insulating alumina barrier and an electrodeposition step. With this technique the resistance of the nanocontacts can be controlled to a certain extent by the indentation parameters. However, Py/$Al_2O_3$/Co (Py=$Ni_{80}Fe_{20}$) nanocontacts did not show any magnetoresistance. This may be due to the deterioration of the alumina layer during indentation or to the electrodeposition process.

In this Letter, we report on a real-time-controlled nanoindentation technique to take a nanocontact on a conductive sample recovered by an insulator, performed using a conducting tip atomic force microscope (CTAFM). This new technique relies on a real-time measurement of the conductance between the tip and a conductive underlying sample, while performing the insulator indentation. Consequently, the exact thickness and mechanical properties of the insulator are not critical as the indentation depth is monitored in real time through the tip-sample resistance. Therefore, this technique allows to stop the indentation with a great precision and to prevent the deterioration of the underlying layers, as evidenced by the observation of magnetoresistance on a tunnel nanojunction fabricated by this new technology.

The CTAFM is based on a Digital Instruments Nanoscope III multimode AFM. This apparatus was modified by Houzé *et al* [12] to perform local resistance measurements in the range of 100 to $10^{12}$ ohms under a bias voltage ranging from 0.1 to 10 V, with 5% accuracy. Conductive tips are provided by Nanosensor. These tips are standard $Si_3N_4$ tips with spring constant ranging from 2 to 50 $N.m^{-1}$ coated by Bore-doped polycrystalline diamond. The macroscopic tip radius is about 100 nm but the diamond crystallites induce a nanoroughness which leads to a local radius of less than 10 nm. During the indentation, the sample is supported by a piezoelectric crystal witch extends the tip vertically upward. The heterostructures were grown on low-conductivity silicon wafers by sputtering. The insulator to be indented is a photoresist layer of approximately 40 nm in thickness. Following the work of Wiesauer *et al* [13], the thin resist layer is obtained by spin coating a Shipley S1805 positive photoresist on the samples at 5500 rpm during 30 s. The thickness of the resist is tuned by diluting the S1805 resist with Shipley EC-solvant. A hard bake step is used to adjust the mechanical properties of the resist [13] and to allow subsequent photolithographic processes without damaging this 40 nm layer.

Figure 1 shows experimental curves obtained while indenting the thin photoresist deposited on a gold surface. We plot the cantilever deflection and the tip-sample resistance as a function of the displacement of the piezoelectric crystal (Δz). We can schematically distinguish 3 steps as the piezoelectric crystal extends:

1) The piezoelectric crystal extends but the tip is not in contact with the resist. The deflection of the cantilever is zero and the tip-sample resistance saturates the capability of the measurement system ($R > 6.10^{11}$ Ω).
2) The tip is indenting the resist. The deflection starts to increase but the tip-sample resistance is still saturated because the tip is isolated from the underlying metallic surface by a too large insulating resist thickness.
3) The tip approaches and finally contacts the metallic surface. The tip-sample resistance exhibits a sharp decrease of about seven orders of magnitude (from $R>6.10^{11}$ Ω to $10^4$ Ω) within a Δz range of about 100 nm.

If the third step of the process is observed more carefully, it appears that it can be subdivided into two regimes: in the first regime, from $6.10^{11}$ to $\sim 10^5$ Ω, the logarithm of the resistance decreases linearly with the piezoelectric crystal extension. In the second regime, from $10^5$ to $\sim 10^4$ Ω, the decrease is much slower (see inset of figure 1). In the first regime, the tip penetration into the resist is proportional to Δz. Therefore, the linear decrease of the logarithm of the resistance with Δz implies that the resistance decreases exponentially with the thickness of the resist located between tip and metallic surface. Thus, the conduction between the tip and the metallic samples occurs by tunneling through the remaining resist. Since 620 nm is the value of Δz needed to indent the whole thickness (40 nm), the resist thickness corresponding to the tunneling regime can be evaluated to 2-3 nm. In the second regime, the resistance variation with Δz becomes much slower. As confirmed by indentation experiments carried out directly on a gold surface, this corresponds to the tip penetration into the gold layer, leading to an increase of the contact surface between the tip and the gold layer.

If one considers the tip penetration into the gold layer to be proportional to $\Delta z$, the contact surface is proportional to square of the penetration depth. As can be seen in the inset of Figure 1, R is indeed proportional to $(\Delta z)^{-2}$ in this regime.

In summary, if we use the tip-sample resistance as the trigger signal to stop the nanoindentation process, we are able to :

1) Stop the indentation process within ~3 nm of remaining resist before the sample surface, or just onto the surface.
2) Stop the indentation process in the underlying sample to create a hole of controlled section.

After indentation, the nanoindented holes are inspected by tapping mode AFM with ultrasharp tips from Olympus having a nominal tip radius of 5 nm. The typical geometry of a hole is presented in Figure 2a. Its profile exhibits a width at half depth of about 35 nm. The bottom lateral dimension is difficult to assess because of the finite tip radius used for imaging, but is definitely smaller than 10 nm (Figure 2b). A nanocontact onto the underlying sample is taken by filling the indented hole with a metal by sputtering, forming the counter-electrode. Finally, this counter-electrode is structured by standard photolithography and ion etching, to allows the measurement of the nanocontacts electrical properties (Figure 3a). It has been verified that structures with non indented resist are insulating (R> several GΩ with bias voltages of several V).

A preliminary experiment carried out to elaborate Au-Au metallic nanocontacts lead to resistance values from 10 Ω to 500Ω. Assuming, roughly, that the quantum resistance of 12.9 kΩ is associated with one atom having an area of about 0.1 $nm^2$, the diameter of the contact is approximately $[10^3/R(\Omega)]^{1/2}$ [14]. A resistance of 500 Ω (respectively 10 Ω) thus reflects a contact with a diameter between 1 and 2 nm (respectively about 10 nm). These values are far below what is achievable with other classical or advanced lithography techniques.

If one wants to define tunnel nanojunctions, areas in the $nm^2$ range will yield resistances of several GΩ. It might therefore be useful to slightly enlarge the holes in order to define junctions with a measurable resistance level. This can be done by a short plasma exposure, which also helps to remove any residual resist. Figure 2c presents the profile of a hole after a 30 seconds 15 W $O_2$ plasma exposure. The flat part of this profile corresponds to the sample surface, which is not etched by the $O_2$ plasma, and has a width of about 10 nm. Finally, by filling the indented holes with a ferromagnetic metal, a nanojunction is created.

In figure 3b, we show a tunnel magnetoresistance (TMR) curve obtained at 300K on a nanojunction fabricated from a $Co/Al_2O_3$ layer. The Co layer is 150 Å thick and the alumina layer is formed by oxidation of a 6 Å Al layer. The counter electrode is a 100 nm thick Co layer. In this nanojunction having a high resistance level of 93 MΩ, a clear TMR signal of 5.1 % is observed. This evidences that during the nanoindentation process, the alumina layer was not destroyed. We have also used this technology to realize the first magnetic tunnel junctions based on half-metallic $Sr_2FeMoO_6$ oxide thin films and to measure spin accumulation effects on clusters. These results will be reported in separate papers.

In summary, we have developed a new technology allowing to fabricate nanocontacts with lateral sizes in the sub-10 nm range. It relies on the electrically-controlled nanoindentation of a thin insulating resist with a conductive tip atomic force microscope, so as to create a nanometric access point onto an underlying conducting structure. The real-time monitoring of the indentation process provides a precise control of the indents depth and potentially of their dimensions. The as-created holes are subsequently filled by metal, and the nanocontact is created. This technique proves to be reliable to produce sub-10 nm structures and especially nanocontacts, either metallic or tunnel-type. Tunnel nanojunctions designed by this technique show a clear magnetoresistance response. The resolution that can be achieved is comparable or better than that obtained by state-of-art electron, ion-beam lithography or even classical AFM lithography. Furthermore, our method allows to control the precise location of the nanocontacts and is compatible with other lithography techniques.

We are very grateful to A. Vaurès for fabrication the multilayers by sputtering, to B. Ba, J. Vanwinsberghe, A. Desroy, M. Le Tacon, M. Le Du and F. Suamaa for technical help and to J.-M. George, H. Jaffrès, and M. Muñoz for helpful discussions. This work has been partially financed by the SPM department of CNRS through an ATIP “Jeune Chercheur”.

References

[1] G. Rubio, N. Agraït and S. Vieira, Phys. Rev. Lett. **76**, 2032 (1996)

[2] A.G.M. Jansen, A.P. van Gelder and P.Wyder, J. Phys.: Condens. Matter **13**, 6073 (1980)

[3] M.V. Tsoi, A.G.M. Jansen, and J.Bass, J. Appl. Phys. **81**, 5530 (1997)

[4] S.M. Rezende, F.M. de Aguiar, M.A. Lucena and A.Azevedo, Phys. Rev. Lett. **84**, 4212 (2000)

[5] J. Grollier, V. Cros, A. Hamzic, J.-M. George, H. Jaffrès, A. Fert, G. Faini, J. Ben Youssef and H. Le Gall, Appl. Phys. Lett. **78**, 3663 (2001).

[6] S. Guéron, M.M. Deshmukh, E.B. Myers, and D.C. Ralph, Phys. Rev. Lett. **83**, 4148 (1999)

[7] R. Deschmidt, G. Faini, V. Cros, A. Fert, F. Petroff and A. Vaurès, Appl. Phys. Lett. **72**, 386 (1998)

[8] M. Viret, S. Berger, M. Gabureac, D. Olligs, I. Petij, J.F. Gregg, F. Ott, G. Francinet, G. Le Goff and C. Fermon, Phys. Rev. B **66**, 220401 (2002).

[9] S.Z. Hua and H.D. Chopra, Phys. Rev. B **67**, 060401 (2003)

[10] N. García, M. Muñoz and Y.W. Zhao, Phys. Rev. Lett. **82**, 2923 (1999)

[11] J. Carrey, K. Bouzehouane, J.-M. George, C. Ceneray, T. Blon, M. Bibes, A. Vaurès, S. Fusil, S. Kenane, L. Vila and L. Piraux, Appl. Phys. Lett. **81**, 760 (2002)

[12] F. Houzé, R. Meyer, O. Schneegans and L. Boyer, Appl. Phys. Lett. **69**, 1975 (1996)

[13] K. Wiesauer and G. Springholz, J. Appl. Phys. **88**, 7289 (2000)

[14] N. García, M. Muñoz, G.G. Qian an H. Rohrer and I.G. Saveliev and Y.W. Zhao, Appl. Phys. Lett. **79**, 4550 (2002).

Figure captions

Figure 1 : Variation of the cantilever deflexion and tip-sample resistance as a function of the displacement of the piezoelectric crystal during the indentation process. The schematics illustrate the relative position of the tip and the sample during the three steps (see text). Inset : detail of the dependence of the resistance on the piezoelectric crystal displacement. Two regimes can be distinguished : for $\Delta z > 920$ nm, log(R) varies linearly with $\Delta z$ which reflects tunnelling conduction ; for $\Delta z < 920$ nm the tip is in contact with the sample surface and R varies as $(\Delta z)^{-2}$ due to the increase of the contact surface. The solid line is a fit to the experimental points in the two regimes.

Figure 2 : (a) Tapping AFM image of an indented hole, obtained with a ultrasharp tip (nominal tip radius ~ 5 nm). Cross-section of two indented holes, without (b) and with (c) post-indentation oxygen plasma etch.

Figure 3 : (a) Schematics of a finalized nanocontact, and of the configuration used to perform electrical measurements. (b) R vs H curve measured at 300K for a $Co/Al_2O_3/Co$ tunnel nanojunction.

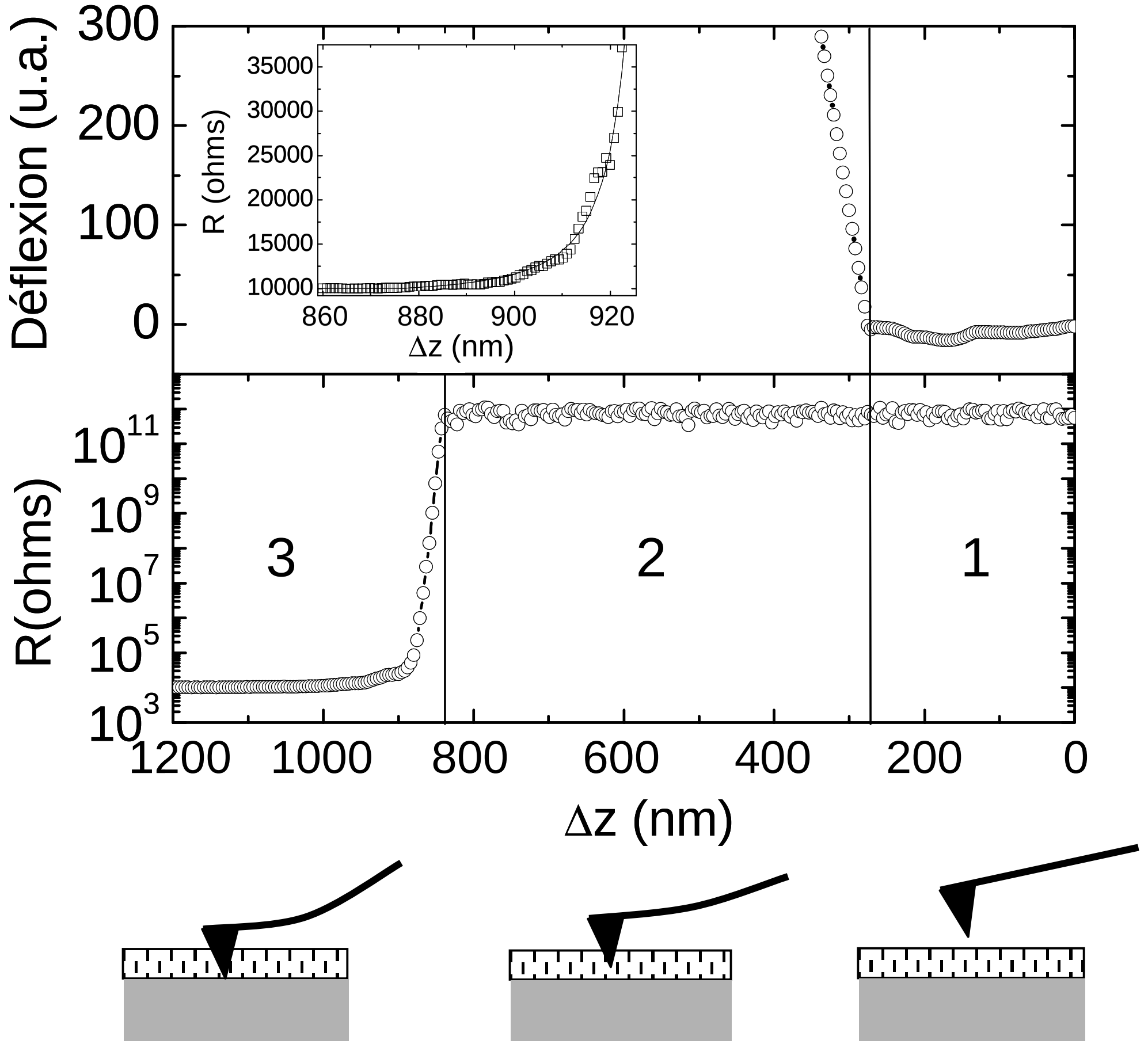

K. Bouzehouane *et al*
Figure 1

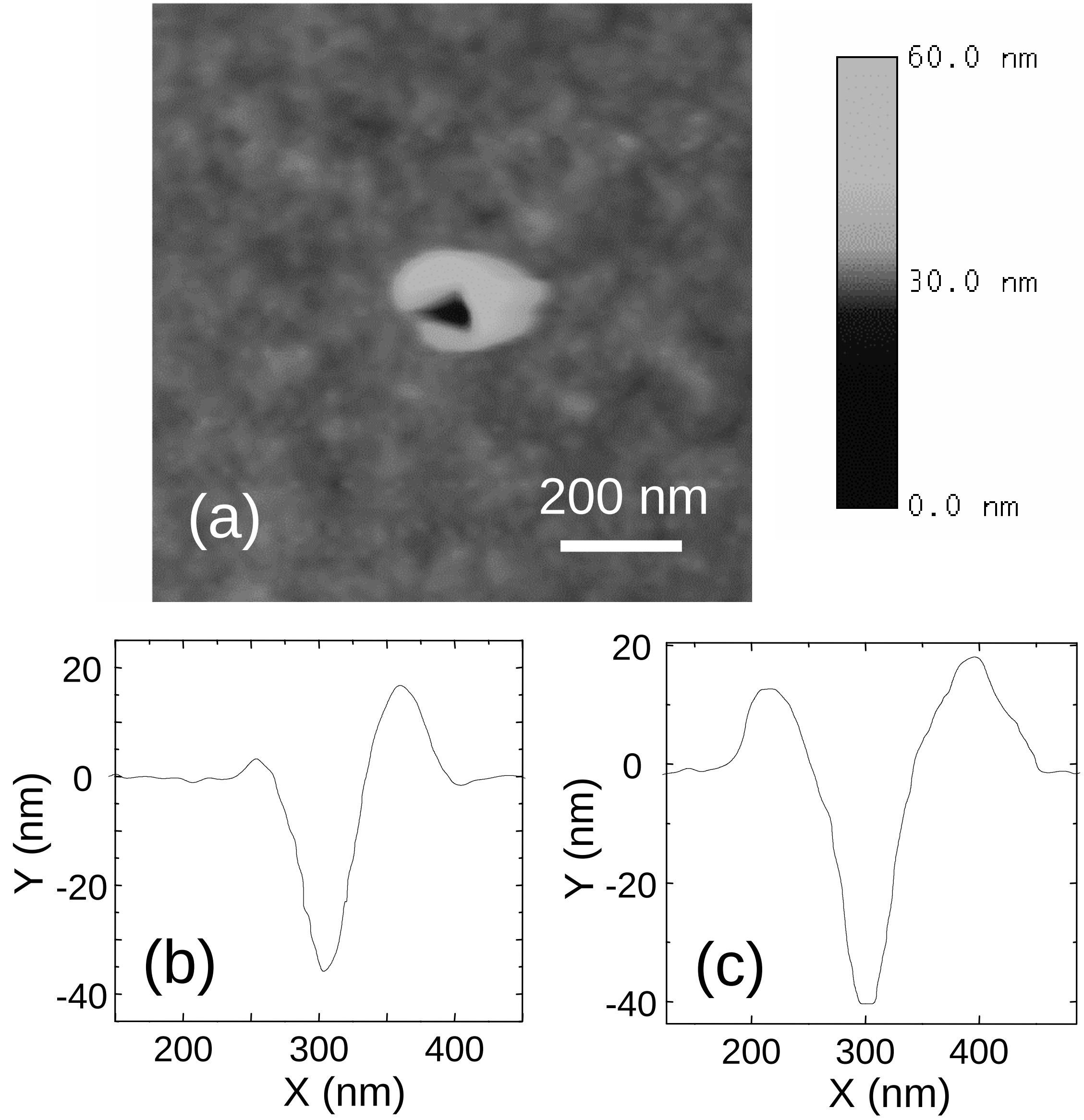

K. Bouzehouane *et al*
Figure 2

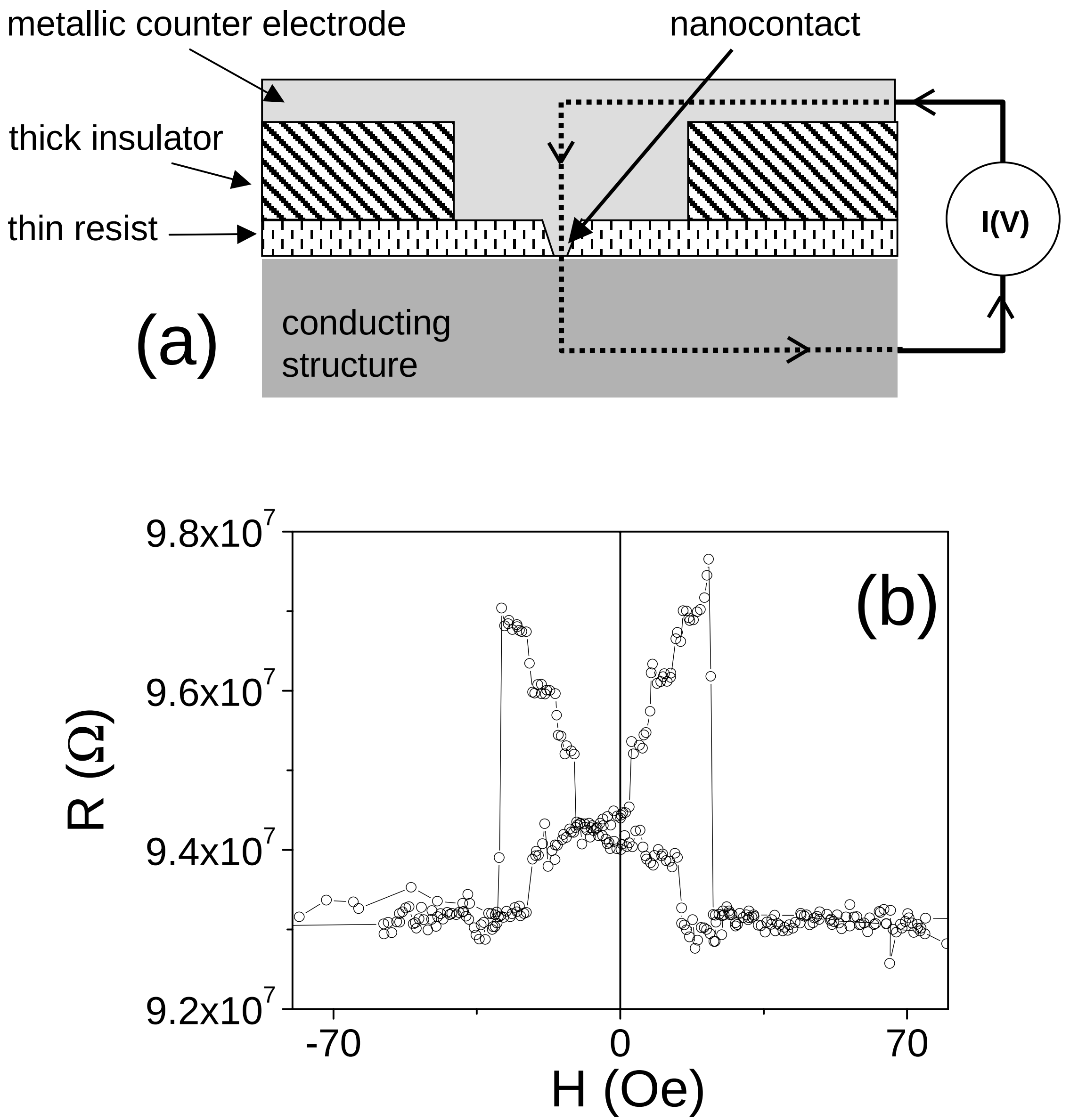

K. Bouzehouane *et al*
Figure 3